\documentclass[conference]{IEEEtran}
\IEEEoverridecommandlockouts
\usepackage{cite}
\usepackage{amsmath,amssymb,amsfonts}
\usepackage{graphicx}
\usepackage{textcomp}
\usepackage{xcolor}
\usepackage{algorithm}
\usepackage{multirow}
\usepackage{algpseudocode}
\usepackage{url}
\usepackage{stfloats} 
\usepackage{lipsum}
\usepackage{caption}
\usepackage{footmisc}

\def\BibTeX{{\rm B\kern-.05em{\sc i\kern-.025em b}\kern-.08em
    T\kern-.1667em\lower.7ex\hbox{E}\kern-.125emX}}

\begin{document}

\title{Uncovering the EEG Temporal Representation of Low-dimensional Object Properties}

\author{
    \IEEEauthorblockN{
        Jiahua Tang\textsuperscript{1, 2, *}, 
        Song Wang\textsuperscript{1},
        Jiachen Zou\textsuperscript{1},
        Chen Wei\textsuperscript{1,3,\dag},
        Quanying Liu\textsuperscript{1,\dag}
    }\\
    \IEEEauthorblockA{
        \textsuperscript{1}\textit{Department of Biomedical Engineering, Southern University of Science and Technology, Shenzhen, China}\\ \textsuperscript{2}\textit{School of Computer Science and Technology, Harbin Institute of Technology, Weihai, China} \\
        \textsuperscript{3}\textit{Department of Psychology, University of Birmingham, Birmingham, United Kingdom}\\
        12150103@mail.sustech.edu.cn; liuqy@sustech.edu.cn
    }
}

\renewcommand{\thefootnote}{}

\twocolumn[{%
\renewcommand\twocolumn[1][]{#1}%

\maketitle
\begin{center}
    \includegraphics[width=\textwidth]{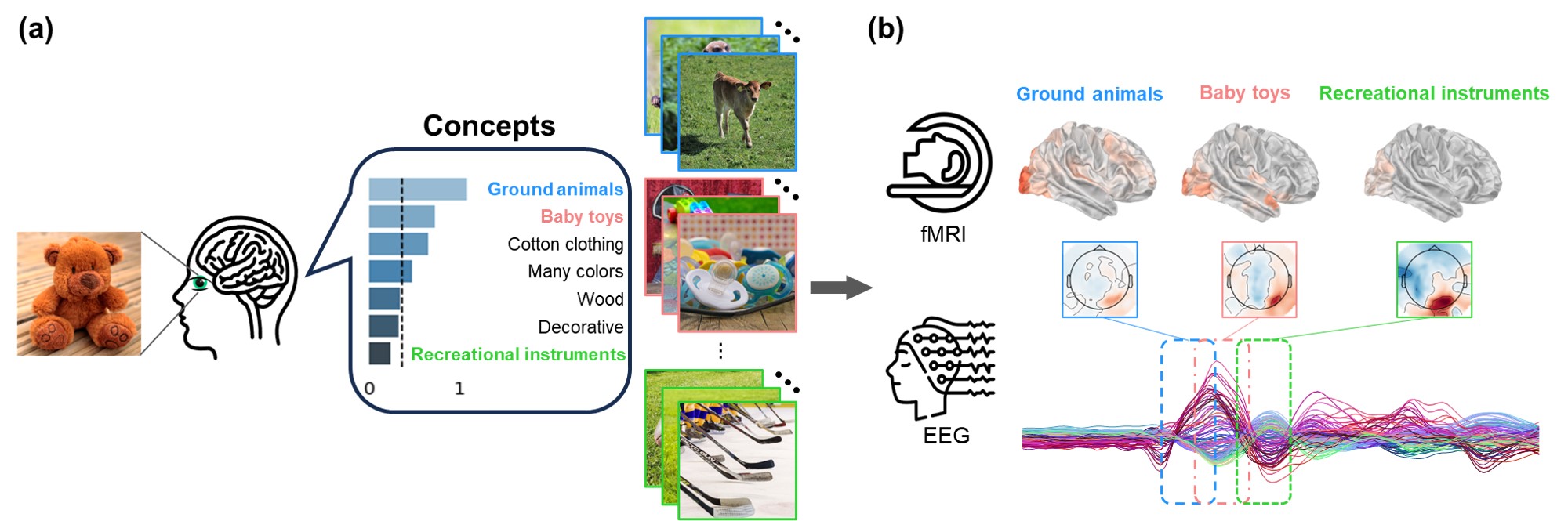}
    \captionof{figure}{\textbf{Motivation of our work.} (Left) A visual stimulus (e.g., a teddy bear) can elicit neural responses associated with conceptual categories (e.g., ground animals, baby toys, and recreational instruments), offering insights into how the brain represents abstract concepts. (Right) EEG and fMRI are two neuroimaging techniques that can be used to record brain activities. While fMRI provides high spatial resolution, facilitating precise localization of conceptual representations in cortical regions, EEG captures the rapid temporal evolution of neural dynamics, making precise positioning challenging.}
    \label{fig:motivation}
\end{center}
}]

\footnotetext{\textsuperscript{\dag} Corresponding author.}
\footnotetext{\textsuperscript{*} Work done while a Visiting Student at Southern University of Science and Technology.}
\footnote{This paper has been accepted by IJCNN2025.}

\begin{abstract}
Understanding how the human brain encodes and processes external visual stimuli has been a fundamental challenge in neuroscience. With advancements in artificial intelligence, sophisticated visual decoding architectures have achieved remarkable success in fMRI research, enabling more precise and fine-grained spatial concept localization. This has provided new tools for exploring the spatial representation of concepts in the brain. However, despite the millisecond-scale temporal resolution of EEG, which offers unparalleled advantages in tracking the dynamic evolution of cognitive processes, the temporal dynamics of neural representations based on EEG remain underexplored. This is primarily due to EEG’s inherently low signal-to-noise ratio and its complex spatiotemporal coupling characteristics. To bridge this research gap, we propose a novel approach that integrates advanced neural decoding algorithms to systematically investigate how low-dimensional object properties are temporally encoded in EEG signals. We are the first to attempt to identify the specificity and prototypical temporal characteristics of concepts within temporal distributions. Our framework not only enhances the interpretability of neural representations but also provides new insights into visual decoding in brain-computer interfaces (BCI).
\end{abstract}
\bstctlcite{IEEEexample:BSTcontrol}


\section{Introduction}

Understanding how the brain encodes visual stimuli has been a central topic in cognitive and neuroscience research, forming the foundation of vision-based BCI. Among human sensory systems, vision serves as the primary channel for external perception, characterized by its complexity and diversity, making it a significant challenge in neural signal decoding. In recent years, breakthroughs in representation learning and generative models have greatly improved the decoding of external visual stimuli from non-invasive neuroimaging modalities such as EEG\cite{fu2024brainvisexploringbridgebrain, FERRANTE2024108701, livisual} and fMRI\cite{xia2024dream, zeng2024controllable, jiang2024mindshotbraindecodingframework}, achieving superior performance in downstream tasks such as retrieval and reconstruction. 
    

\begin{figure}[htbp]
    \centering
    \includegraphics[width=0.4\textwidth]{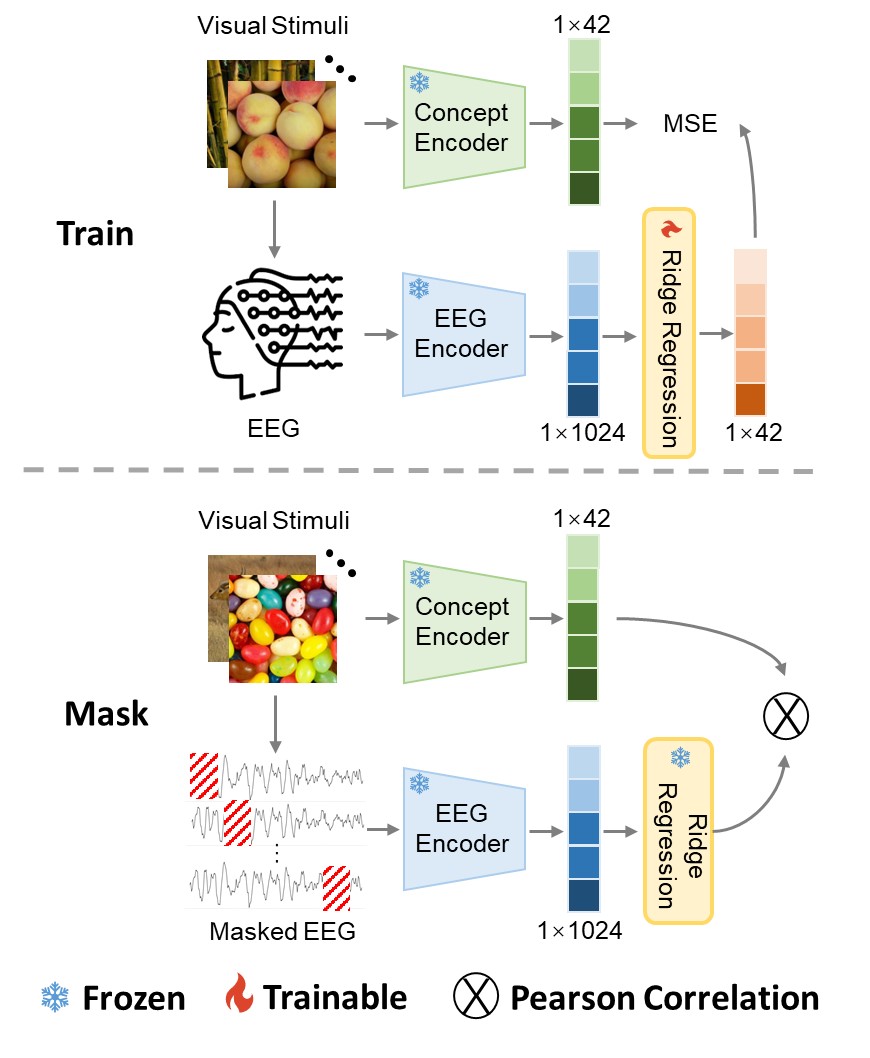}
    \caption{\textbf{Framework for Dynamic Concept Attribution}. (Top) Training phase: The Concept and EEG encoders extract embeddings for visual stimuli and EEG signals, respectively. A ridge regression model is then trained to map the EEG representations to the latent concept space. (Bottom) Temporal Masking: EEG signals with masked time points are used to predict concept activation, which is quantified using the Pearson correlation to assess the impact of different time points on concepts.} 
    \label{fig:framework}
\end{figure}

These advancements in visual neural decoding have significantly improved the performance of BCI tasks and paved the way for further exploration of human brain visual representations. Specifically, recent studies have proposed leveraging fMRI’s spatially decoupled signals for concept localization\cite{beauchamp2007grounding, harpaintner2020grounding, luo2024brainmappingdensefeatures}. For example, BrainSCUBA \cite{luo2024brainscuba} employed data-driven language-vision alignment to uncover more fine-grained semantic representations in the human visual cortex. Similarly, \cite{shen2024neuro} utilized post-perturbation reconstruction to map semantic selectivity in fMRI activity. However, these studies rely on fMRI’s inherent spatial decoupling property, where visual concepts exhibit clear spatial distributions within the cortex. When attempting to investigate the dynamic processing of visual information in the brain using EEG signals, the strong temporal coupling of EEG leads to the mixing of neural dynamics across different brain regions. This fundamental difference makes it challenging to directly apply concept localization methods developed for fMRI to EEG signals. Consequently, it motivates us to explore novel concept localization approaches tailored specifically for the EEG modality, as illustrated in Fig. \ref{fig:motivation}.


To overcome the coupling characteristics and low signal-to-noise ratio of EEG signals and accurately identify the representation specificity of different concepts in EEG, selecting appropriate visual concepts for localization is crucial. First, unlike fMRI, fine-grained concepts are difficult to express in the low-SNR EEG signals. Second, due to the brain’s information processing mechanisms, different levels of information (such as visual and semantic features) exhibit distinct temporal dynamics, whereas information within the same level (such as category information) is harder to differentiate based on temporal dynamics alone. Therefore, the selected concepts need to meet both coarse granularity and multi-level requirements.

To address this, scientists have proposed a series of decision-making tasks based on visual stimuli \cite{murphy1985role, medin1993respects, roads2024modeling} to explore more abstract human concepts of visual objects. Notably, \cite{teichmann2023dynamic} introduced a novel behaviorally related framework that directly examines how behavior-derived object dimensions are reflected in the dynamic object representations of the human brain, revealing the temporal dynamics of object processing. Inspired by these studies, we adopt behavior-based low-dimensional visual concepts. These concepts are widely present across all images and exhibit a temporal sequence in the brain’s visual information processing (e.g., visual and semantic features). Therefore, we will primarily discuss their temporal dynamics.


Our framework integrates concept mapping, dynamic temporal masking, and patch attribution to decode behavior-driven low-dimensional object properties from EEG signals. As shown in the framework diagram in Fig. \ref{fig:framework}, by training a cross-modal encoder to align EEG temporal dynamics with visual concept embeddings, we develop a temporal masking mechanism that identifies critical time windows for concept activation. This method not only addresses EEG’s spatiotemporal coupling challenges but also enables precise localization of neurocognitive processes. Its applications extend to adaptive brain-computer interfaces (BCI), where temporally resolved concept representations enhance real-time decoding, and to cognitive neuroscience, providing a tool to map the hierarchical progression of visual-to-semantic processing in the human brain.

Our principal contributions are threefold:

(1) We propose a novel method integrating dynamic temporal masking and patch attribution to uncover temporally specific concept representations in EEG signals, addressing their inherent spatiotemporal coupling and low signal-to-noise ratio.

(2) By establishing a cross-modal analytical framework through behaviorally defined low-dimensional concept spaces, we bridge spatial and temporal neural representations, enabling interpretable decoding of EEG dynamics.

(3) We systematically identify hierarchical temporal patterns in neural encoding through clustering analysis, revealing distinct concept clusters (e.g., early visual processing, multi-feature integration, and late semantic representation) that reflect the brain's staged information processing mechanism.

\section{Related works}


\textbf{Behavioral Based Low-dimension Visual Embedding.} In recent years, behavior-based low-dimensional visual representation methods have gained significant attention, particularly in the fields of neuroscience and computer vision. Previous studies \cite{hebart2019things, hebart2020revealing, zheng2019revealing, muttenthaler2022vice} collected human similarity judgment data for a naturalistic dataset of 26,107 object images through large-scale behavioral experiments, and decoded the representation dimensions of images by either optimizing the representations of individual objects or using deep neural networks (DNNs) to predict human behavior, thereby compressing high-dimensional visual information into low-dimensional visual representations. Furthermore, recent studies \cite{wei2024cocog2, wei2024cocog} have implemented controllable image generation based on low-dimensional representations by fitting behavior-decoded visual representations; \cite{li2024visual} showed these low-dimensional representations can also be decoded by EEG signals with high consistency. These studies demonstrate the effectiveness of low-dimensional visual representations in computer vision and neural decoding research.

\section{Method}
In this section, we describe our model in detail, which combines concept mapping, dynamic masking, and temporal patches attributing.
\subsection{Concept Mapping}\label{mapping}
Consider the paired \{EEG, image\} dataset defined as $\Omega = \{ (X_i , I_i) \}_{i=1}^{n}$, where each pair consists of a multi-channel EEG signal $X_i \in \mathbb{R}^{C \times T}$ and its corresponding visual stimulus $I_i$. Here, $C$ denotes the number of EEG electrodes (channels) and $T$ represents the total timestamps. The variable $n$ represents the total number of the paired data. We adapt the architecture of \cite{li2024visual} as the EEG encoder, denoted as $f(\cdot)$, to obtain the EEG embedding $E_e = f(X_i)$. The EEG encoder \( f(\cdot) \) is typically based on the channel-wise Transformer encoder, Temporal-Spatial convolution and multilayer perceptron (MLP) architecture. We also adapt the architecture of \cite{wei2024cocog} as the concept encoder, denoted as \( g(\cdot) \), to obtain the concept embedding \( E_c = g(I_i) \). Specifically, we first extract the CLIP embedding from the image \( I_i \), and then use a concept projector, which maps the CLIP embedding to the concept space, resulting in the concept embedding \( E_c \). Let $E_e \in \mathbb{R}^{F_e}$ and $E_c \in \mathbb{R}^{F_c}$. $F_e$ and $F_c$ represent the dimensions of the EEG embedding and the concept embedding, respectively.

To map the EEG embedding $E_e$ to the concept embedding $E_c$, we employ Ridge regression:

\begin{equation}
E_c \approx W E_e + b
\end{equation}

where $W \in \mathbb{R}^{F_c \times F_e}$ is the weight matrix and $b \in \mathbb{R}^{F_c}$ is the bias vector.

To mitigate overfitting risks in our regression model, we augment the objective function with an $L_2$ regularization term, resulting in the following Ridge regression formulation:

\begin{equation}
\mathcal{L} = \frac{1}{n} \sum_{i=1}^{n} \| E_c^i - (W E_e^i + b) \|^2 + \lambda \|W\|^2
\end{equation}

where $\lambda$ is the regularization hyperparameter that controls the trade-off between minimizing reconstruction error and penalizing large weights.

\subsection{Dynamic Masking}
To investigate the temporal contribution of different EEG segments to concept embedding activation, we introduce a dynamic time masking strategy. By selectively masking different time intervals in the EEG signals, we analyze how these masked representations influence the mapped concept embedding.

As defined in \ref{mapping}, $T$ is the total number of timestamps in EEG and a masked EEG sequence $\tilde{X}_i^{(t_k, L)}$ can be defined as:

\begin{equation}
\tilde{X}_i^{(t_k, L)}[:, t_k:t_k+L] = 0
\end{equation}

where $(t_k, L)$ denotes the starting time index $t_k$ and mask length $L$. The masked EEG embedding is obtained as:

\begin{equation}
\tilde{E}_e^{(t_k, L)} = f(\tilde{X}_i^{(t_k, L)})
\end{equation}

The corresponding concept embedding is then predicted using the trained regression model:

\begin{equation}
\tilde{E}_c^{(t_k, L)} = W \tilde{E}_e^{(t_k, L)} + b
\end{equation}

To quantify the impact of masked EEG segments on concept embeddings, we compute the Pearson correlation coefficient between the original and mask computed concept embeddings.

Given the original concept embedding $E_c$, the Pearson correlation coefficient is defined as:

\begin{equation}
\rho(t_k, L) = \frac{\sum_{j=1}^{F_c} (E_c^j - \bar{E}_c) (\tilde{E}_c^{j, (t_k, L)} - \bar{\tilde{E}}_c^{(t_k, L)})}
{\sqrt{\sum_{j=1}^{F_c} (E_c^j - \bar{E}_c)^2} \sqrt{\sum_{j=1}^{F_c} (\tilde{E}_c^{j, (t_k, L)} - \bar{\tilde{E}}_c^{(t_k, L)})^2}}
\end{equation}

where $\rho(t_k, L) \in [-1,1]$. Higher values indicate weaker relevance between the masked and original embeddings, while lower values indicate a greater impact of the masked EEG segment.

\subsection{Temporal Patches Attributing}
To further analyze the temporal structure of EEG contributions to concept embedding, we employ Dynamic Time Warping (DTW)\cite{sakoe1978dynamic}.

Let $\rho_i(t_k, L)$ denote the Pearson correlation sequence for the $i$-th EEG trial, where each sequence represents the impact of different masked segments on concept embedding. We compute the DTW distance between any two sequences $\rho_i(t_k, L)$ and $\rho_j(t_k, L)$ as:

\begin{equation}
D_{\text{DTW}}(i, j) = \text{DTW}(\rho_i(t_k, L), \rho_j(t_k, L))
\end{equation}

where $\text{DTW}(\cdot, \cdot)$ is the dynamic time warping function that computes the optimal alignment cost between two sequences.

Next, we construct a distance matrix $D$ such that:

\begin{equation}
D_{ij} = D_{\text{DTW}}(i, j)
\end{equation}

We then apply hierarchical clustering with linkage on $D$ to identify distinct groups of EEG contribution patterns. The final clusters are obtained using a predefined number of clusters $K$:

\begin{equation}
C = \text{Cluster}(D, K)
\end{equation}

where $C$ represents the cluster assignment of each EEG trial. This clustering process allows us to identify common EEG temporal dynamics that contribute to specific concept activations.

\begin{figure*}[t]
    \centering
    \includegraphics[width=1.0\textwidth]{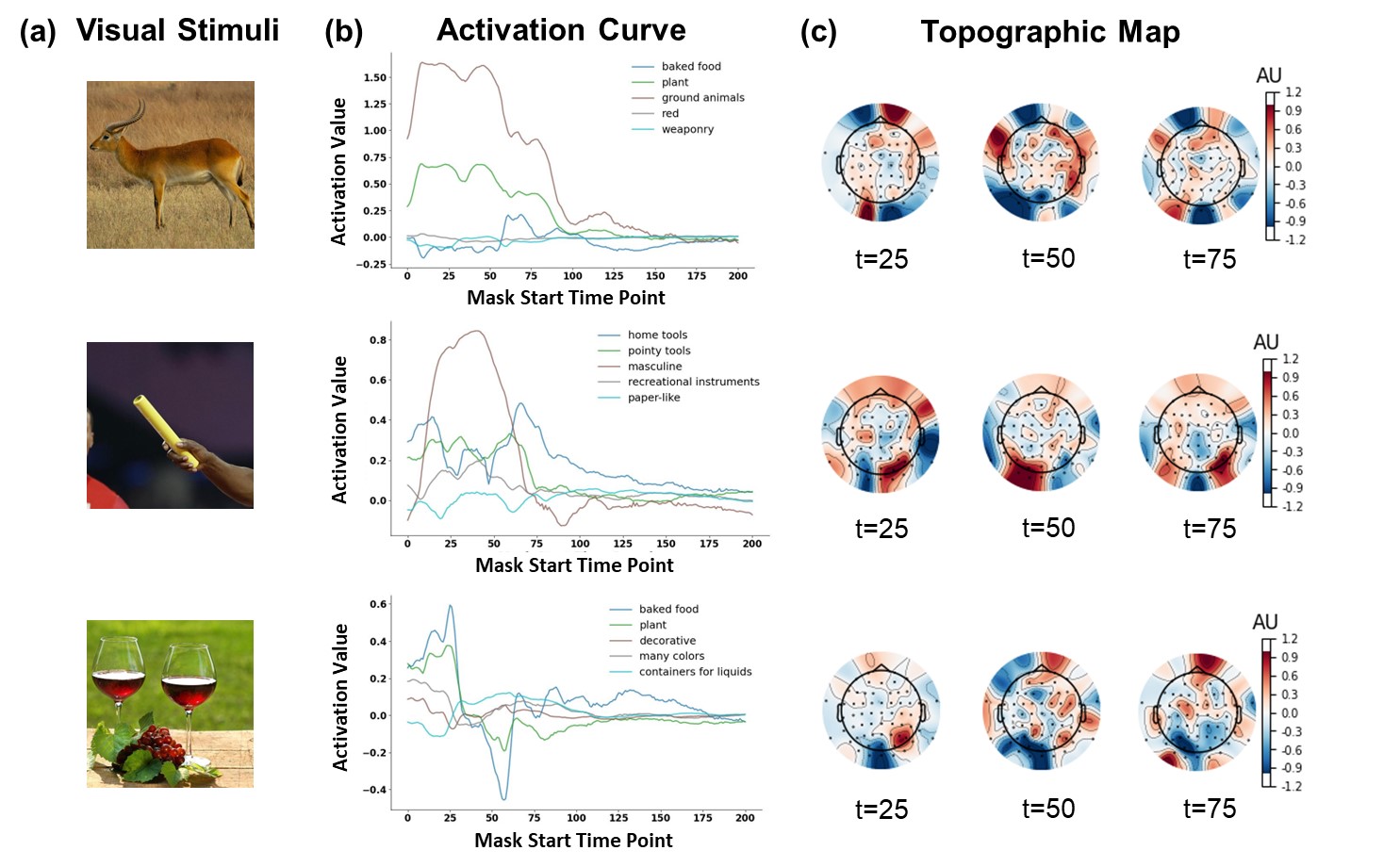}
    \caption{\textbf{Temporal Impact of EEG Signals on Concept Activation.} (a) Visual object presented to the subject. (b) For the top five concepts most strongly activated by the image, we apply temporal masking to the EEG signals at different start times and then calculate the difference in predicted concept activation values before and after masking. (c) Topographic maps of EEG signals at $ t = 25,  50, 75$ ms.} 
    \label{fig:concept}
\end{figure*}

\section{Experiments}
\subsection{Experiments Setup}
We conducted experiments using the THINGS-EEG dataset \cite{grootswagers2022human}, which follows the rapid serial visual presentation (RSVP) paradigm, where each image is presented for 50 ms, followed by a 50 ms blank screen. Each EEG data is recorded for 1 second, using a 64-channel system with a 1,000 Hz sampling rate. After preprocessing, the data is downsampled to 250 Hz. The training set includes 1,654 distinct categories, each containing 10 images, with each image presented 4 times, resulting in a total of 66,160 EEG samples. The test set consists of 200 categories, each represented by a single image repeated 80 times, yielding 16,000 EEG samples. These images represent a wide range of visual concepts, ensuring diversity in the stimuli presented to participants. Notably, all categories in the test set are entirely distinct from those in the training set.

All experiments were conducted on a single NVIDIA 4090 GPU. For the training process of the EEG encoder, the EEG signals were transformed into a 1×1024 latent representation using the EEG encoder, which was trained according to the parameters specified in \cite{li2024visual}. The concept encoder, on the other hand, utilized pre-trained weights from \cite{wei2024cocog} to transform the corresponding images into a 1×42 latent space. For in-subject model training, we utilized Ridge regression referred to \cite{teichmann2023dynamic} with a regularization parameter $\lambda = 0.5$ to map the EEG embedding to the concept embedding on a dataset of 66,160 EEG samples. For the test data, to improve the signal-to-noise ratio and enhance the quality of the EEG data, we averaged the 80 repetitions for each image. This averaging process reduces the variability and noise inherent in individual EEG recordings, resulting in cleaner and more reliable data. 

\subsection{Concept Mapping Result in Time-series}
To investigate the temporal representation of distinct concepts in EEG signals, we employed a dynamic masking approach, systematically masking continuous 50-timepoint segments of the original EEG signal within the test dataset, starting from timepoint 1 to 200. These masked EEG signals were then passed through a pre-trained EEG encoder to obtain the EEG embedding, which was subsequently used with a pre-trained regression model to predict the corresponding activation values for each concept. We focused on the top-k most strongly activated concepts and examined how their predicted values changed with different masking start times. These changes reflect the importance of specific time segments for representing each concept. We selected three different subjects (sub-05, sub-08, and sub-10) and three distinct visual stimuli (shown in Figure \ref{fig:concept} (a)) as illustrative examples. As depicted in Figure \ref{fig:concept} (b), our findings reveal that different concepts exhibit temporally specific activation patterns. For example, in the second row, a clear chronological order of concept activation is observed (e.g. home tools and paper-like). Additionally, comparing the first and third rows, the concept 'plant' appears across both subjects and stimuli, but its activation timing varies, highlighting both consistency in concept activation and variability in temporal dynamics.

\begin{figure}
    \centering
    \includegraphics[width=0.5\textwidth]{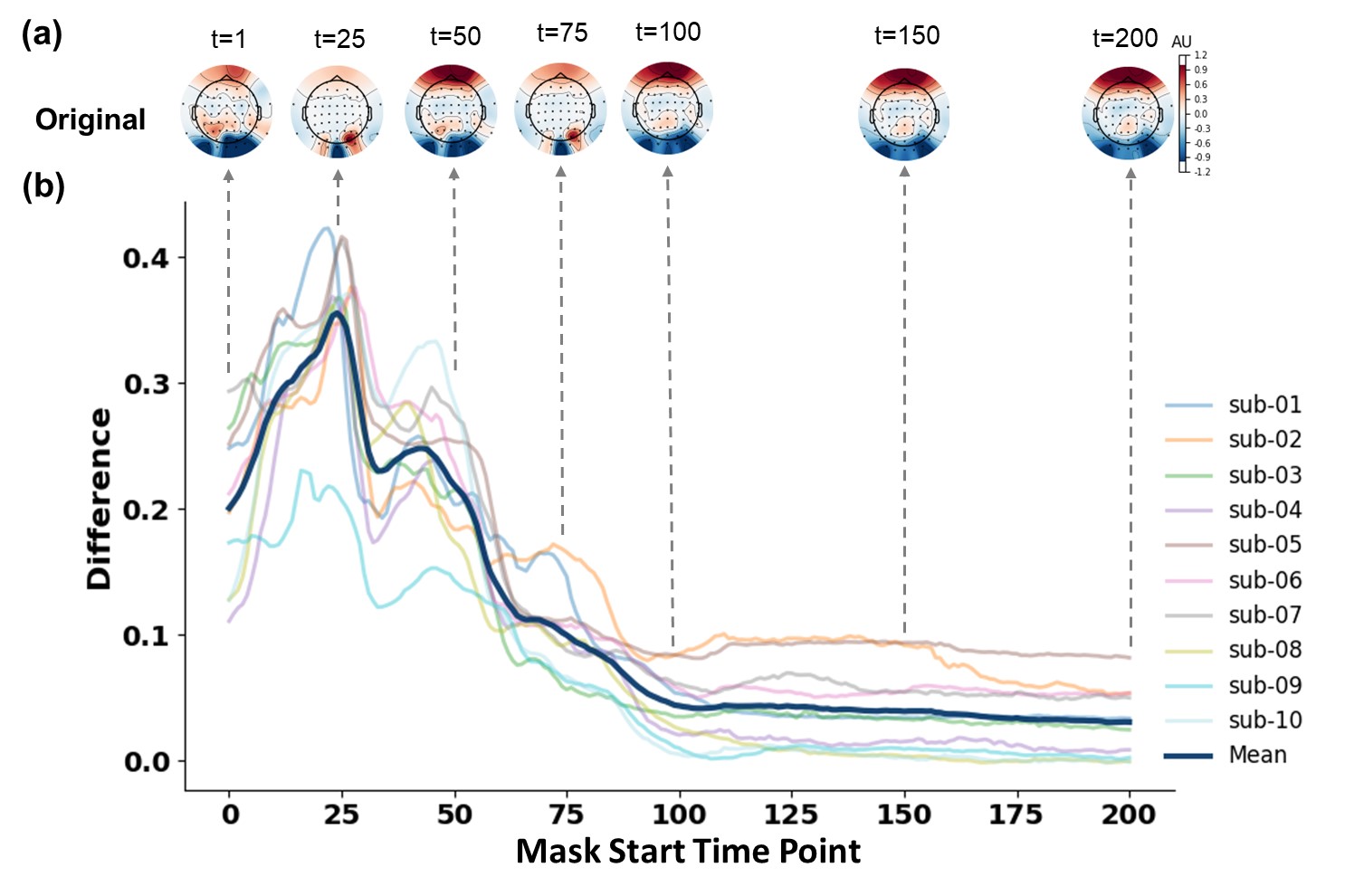}
    \caption{\textbf{Impact of Temporal Masking on Concept Activation Across Subjects.} (a) Averaged topographic maps of the original EEG signals, plotted at 25 ms intervals within the 0-200 ms window.  (b) For each subject's EEG signals, temporal masking is applied at different start times. The curves for sub-01 to sub-10 represent the average difference in Pearson correlation coefficients between the predicted concept activation values and the true values before and after masking, calculated across all samples of each subject. The 'Mean' curve represents the average of these differences across all subjects.} 
    \label{fig:erp-across}
\end{figure}

\begin{figure*}[h]
    \centering
    \includegraphics[width=1.0\textwidth]{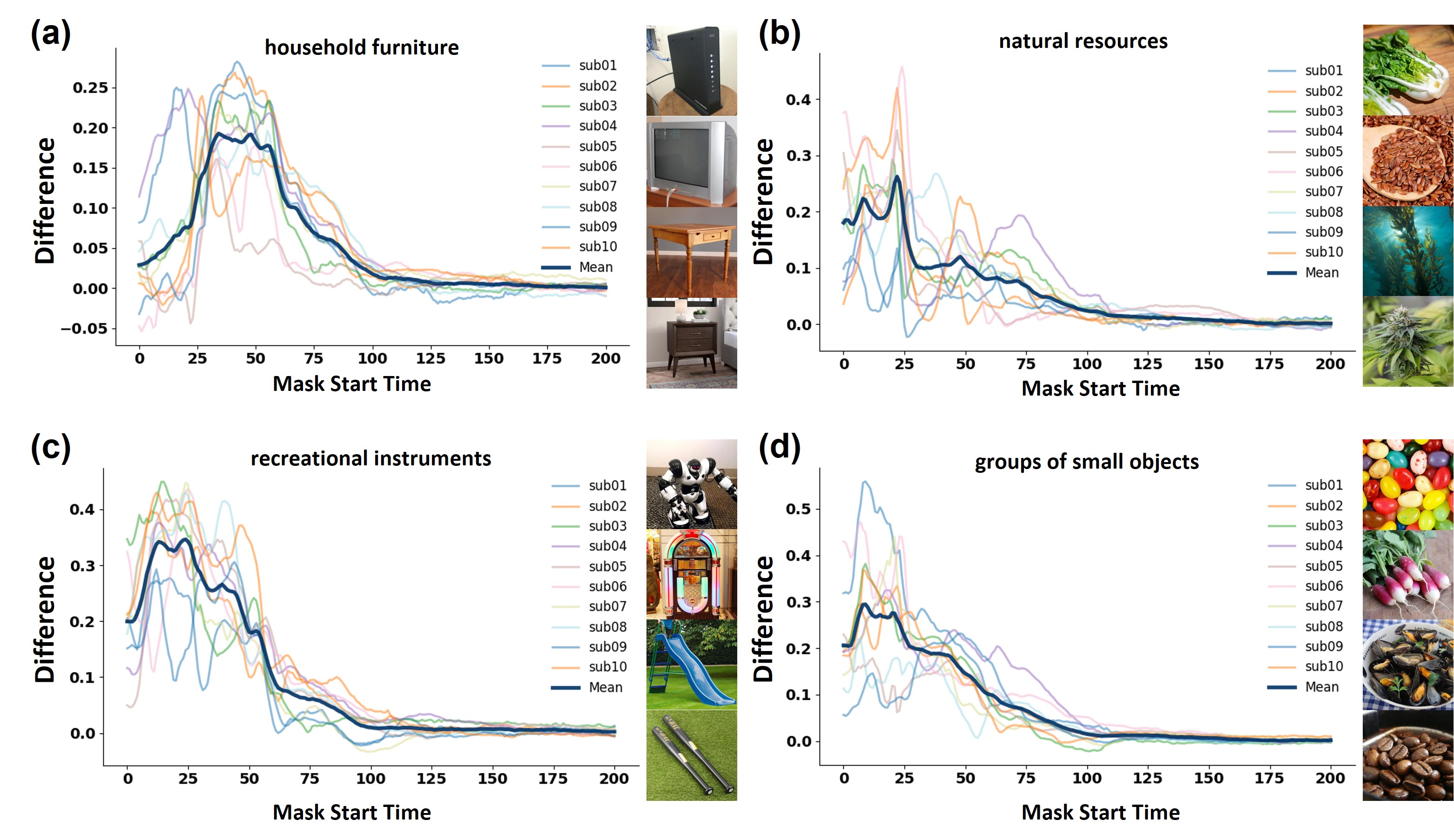}
    \caption{\textbf{Impact of Temporal Masking on Specific Concept Activation.} For four example concepts, the EEG signals corresponding to each concept are calculated for each subject and temporally masked at different start times. On the left side of each concept subplot, the difference in Pearson correlation coefficients between the predicted and true concept activation values before and after masking is shown for each subject, along with the average difference. On the right side, the example image with the highest activation for that concept, as determined from EEG signals, is displayed.} 
    \label{fig:erp-sepe}
\end{figure*}

\subsection{Neuro-related Temporal Schema}
To rigorously analyze the influence of specific EEG signals on concept activation, we computed the Pearson correlation between the predicted activation values derived from the original EEG signals and those obtained from dynamically masked EEG signals. In Figure \ref{fig:erp-across} (a), we present brain topomaps generated from the original EEG signals, which align with well-established temporal patterns of neural responses to visual stimuli. These topomaps offer spatial insight into neural activity and confirm prior findings regarding critical time windows for visual processing. Figure \ref{fig:erp-across} (b) illustrates the difference in Pearson correlation values (original minus masked) across ten subjects and 200 visual stimuli in the test dataset. The resulting curves reveal notable inter-subject variability in both the timing and magnitude of activation, highlighting the inherent heterogeneity in neural responses across individuals. However, by aggregating the data, we identified a consistent pattern characterized by two distinct periods of concentrated visual activation: the first occurring between the mask start point 0 to 25 and the second around 30 to 60 mask start point. This suggests common temporal windows across subjects where visual stimuli significantly impact neural activation.

To visualize concept-specific neural dynamics more intuitively, we introduced a neuro-temporal schema. In Figures \ref{fig:erp-sepe} (a)-(d), we present the Pearson correlation results for four representative concepts from a set of 42. Taking 'household furniture' as an instance,  our analysis reveals that this concept is most strongly correlated with EEG signal segments masked between 25 and 75 timepoints. Additionally, our results demonstrate that the temporal representation of different concepts follows a sequential order, with certain concepts being activated earlier or later than others, reflecting the hierarchical nature of neural processing. Furthermore, we also observed notable inter-subject variability in the precise timing of activation, emphasizing the role of individual differences. To deepen our understanding, we also identified the specific EEG signal patterns that maximally activate the concept and listed the corresponding visual objects, which are displayed alongside the results. Our neuro-related temporal schema provides a powerful way to highlight the temporal dynamics of concept-specific neural activation and also bridge the gap between neural activity and elicited concepts.

\subsection{Prototypical Temporal Characteristics and Clustering}
\begin{figure*}[t]
    \centering
    \includegraphics[width=1\textwidth]{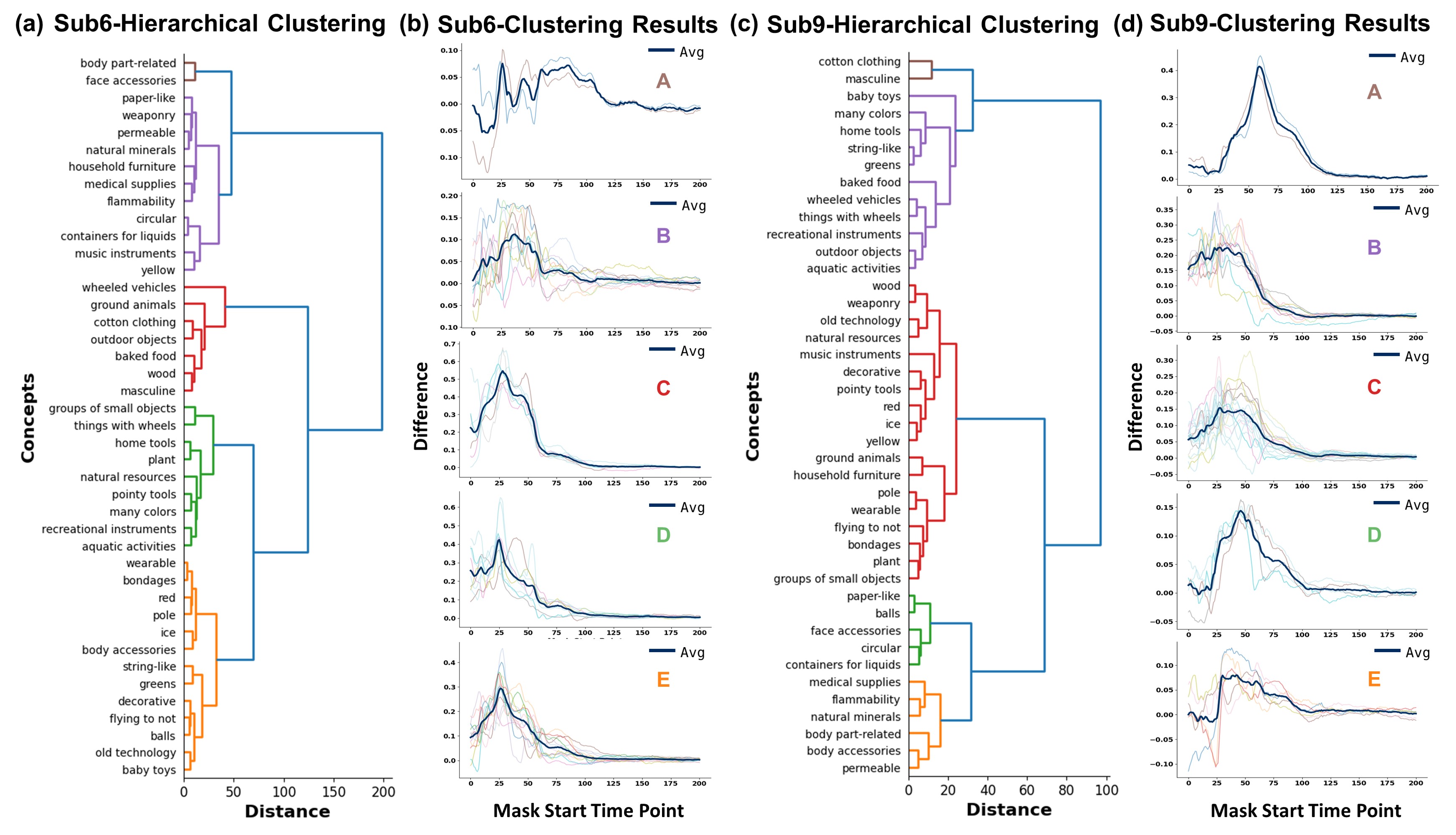}
    \caption{\textbf{Cluster Analysis of Concept Activation.} (a) Hierarchical clustering from subject 9 of the 42 concepts, resulting in 5 clusters.  (b) For each cluster (A–E) in subject 9, the plot shows the difference in Pearson correlation coefficients between the predicted and true activation values before and after masking for each concept within a cluster, along with the average difference across all concepts. (c) Hierarchical clustering from subject 6. (d) Clustered Pearson correlation coefficients from subject 6.} 
    \label{fig:clsuter}
\end{figure*}

To quantify the similarity of timecourse shapes across concepts, we used dynamic-time warping (DTW),  a method that measures the similarity between two time series by calculating the degree of warping required to align one series with the other. By applying DTW to our Pearson correlation data, we generated a distance matrix and subsequently performed hierarchical clustering on this matrix. We selected subjects 6 and 9 as illustrative examples. As shown in Figure \ref{fig:clsuter}, we set the number of clusters to five and present the clustering results for subject 6 in (a) and (b), and for subject 9 in (c) and (d). 

The clustering revealed distinct concept groupings based on temporal activation dynamics. For subject 9, the brown cluster A, containing concepts like 'cotton clothing' and 'masculine', separated earliest from the others. The purple cluster B included a mix of indoor items (e.g., baby toys, baked food) and outdoor objects (e.g., wheeled vehicles, things with wheels). The red cluster C grouped color-related and natural concepts (e.g., red, yellow, wood, ground animals), while the green cluster D focused on human-made items (e.g., face accessories, containers for liquids). Finally, the orange cluster E was dominated by body-related concepts. In Figure \ref{fig:clsuter} (d), we visualize the Pearson correlation patterns from subject 9 within each cluster to examine their prototypical temporal characteristics. Notably, clusters B through E generally exhibit earlier activation compared to cluster A and clusters A through C have stronger activation values. Further comparing the results of the two subjects, we can find that there is a high probability of of being clustered together with human-related or face-related.

\section{Conclusion}

This study introduces a novel framework for analyzing the temporal dynamics of visual concept encoding in EEG signals, aiming to understand how low-dimensional visual properties are encoded over time in the brain. By utilizing advanced concept mapping and dynamic time-masking techniques, our experiments reveal distinct temporal phases in which different object properties exhibit strong neural activation. These findings highlight EEG's ability to capture rapidly evolving representations, suggesting that temporal specificity in EEG may parallel spatial selectivity observed in fMRI-based localization studies. This temporal perspective is crucial for a more granular understanding of how the brain processes visual stimuli in real-time, enhancing the interpretability and applicability of EEG-based decoding frameworks in dynamic cognitive tasks.

We examined the temporal influence of EEG signals on concept activation by masking different time segments of the EEG signal. The results demonstrate that concept activations exhibit clear temporal specificity, with distinct activation patterns across time for different concepts. This dynamic sequence of object feature processing suggests that certain concepts, particularly those involving complex or abstract semantic features, may require a specific sequence of neural activation over time. This finding aligns with studies of human visual pathways, where visual stimuli undergo spatial attention shifts followed by semantic extraction, with neural pathways sequentially reflecting the understanding and representation of concepts\cite{fernandino2022decoding,bausch2021concept}.

We also observed variability in temporal contributions to concept activation across subjects, while overall patterns shared commonalities. This underscores the robustness of our approach, which captures not only universal brain responses but also individual differences in temporal concept representation. The variability in temporal dynamics emphasizes the necessity of considering individual differences in EEG-based decoding models, reflecting the balance between group commonalities and individual variability in visual processing.

Hierarchical clustering of concept activation patterns based on temporal dynamics further revealed the existence of prototypical temporal structures in how the brain encodes visual concepts. These clusters suggest that the brain organizes concept representations into distinct, recurrent temporal patterns, which may reflect fundamental principles of brain organization. These findings are consistent with studies linking specific oscillatory brain waves to brain activity\cite{shine2016dynamics,staresina2024coupled}. Further refinement of clustering methods could reveal more subtle temporal features underlying complex cognitive processes such as memory retrieval, decision-making, and perceptual learning.

Despite promising results, several limitations of the current study must be acknowledged. First, the analysis relies on a specific dataset , which, while diverse, may not fully capture the range of naturalistic visual stimuli and real-world cognitive tasks. Future work could extend this framework by using more ecologically valid datasets, such as dynamic or natural scenes. Additionally, although EEG offers high temporal resolution, its spatial resolution remains limited. Future research may consider techniques like source imaging to derive full-brain spatiotemporal dynamics from EEG data\cite{wang2024advancing,wei2021edge}, improving the spatial accuracy of neural activity interpretation.

Furthermore, while our study focused on low-dimensional object properties, higher-level conceptual processing and abstract semantic representations remain largely unexplored. Future work could expand our framework to incorporate more complex, hierarchical semantic categories, providing deeper insights into how abstract cognitive processes evolve over time in the brain. This could involve studying how different categories of knowledge are temporally encoded within neural networks.

Finally, although the temporal structure of concept activation revealed by clustering analysis is insightful, further exploration is needed to understand how these temporal patterns relate to broader cognitive functions, such as attention, memory, and decision-making. Future studies could manipulate these cognitive functions experimentally to examine their impact on the temporal dynamics of concept activation, thereby enhancing our understanding of the interaction between cognitive processes and their neural representations.

In conclusion, this work presents a powerful framework for analyzing the temporal dynamics of visual concept encoding in EEG signals. By providing both temporal resolution and interpretability, our method opens new avenues for EEG-based research in cognitive neuroscience, particularly in real-time brain-computer interface applications. The ability to track how concepts evolve over time in the brain holds promise for advancing our understanding of dynamic cognition and for developing more sophisticated BCI systems that are responsive to cognitive states in real-time. Furthermore, by uncovering the prototypical temporal structures underlying concept activation, this study lays the foundation for future research into neural organization and provides potential pathways for improving clinical and computational applications of EEG.

\section{Acknowledgment}
This work was supported by the National Natural Science Foundation of China (62472206), Shenzhen Science and Technology Innovation Committee (2022410129, KJZD20230923115221044, KCXFZ20201221173400001), Guangdong Provincial Key Laboratory of Advanced Biomaterials (2022B1212010003), and the Center for Computational Science and Engineering at Southern University of Science and Technology.
\bibliographystyle{IEEEtran}
\bibliography{main}

\end{document}